\documentclass[conference]{IEEEtran}%
\usepackage{amsmath}
\usepackage{graphicx}
\usepackage{amsfonts}
\usepackage{algorithm, algorithmicx, algpseudocode}

\begin{document}
\title{Message passing resource allocation for the uplink of multicarrier systems}
\author{\IEEEauthorblockN{Andrea Abrardo\IEEEauthorrefmark{1}, Paolo Detti\IEEEauthorrefmark{1}, Marco Moretti\IEEEauthorrefmark{2}}
\IEEEauthorblockA{\IEEEauthorrefmark{1}Dipartimento di Ingegneria dell'Informazione,
University of Siena, Italy}
    \IEEEauthorblockA{\IEEEauthorrefmark{2}Dipartimento di Ingegneria dell'Informazione, University of
Pisa, Italy} }
\maketitle

\begin{abstract}
We propose a novel distributed resource allocation scheme for the
up-link of a cellular multi-carrier system based on the message
passing (MP) algorithm. In the proposed approach each transmitter
iteratively sends and receives information messages to/from the base
station with the goal of achieving an optimal resource allocation
strategy. The exchanged messages are the solution of small
distributed allocation problems. To reduce the computational load,
the MP problems at the terminals follow a dynamic programming
formulation. The advantage of the proposed scheme is that it
distributes the computational effort among all the transmitters in
the cell and it does not require the presence of a central
controller that takes all the decisions. Numerical results show that
the proposed approach is an excellent solution to the resource
allocation problem for cellular multi-carrier systems.
\end{abstract}

\section{Introduction}
Orthogonal Frequency Division (OFDM) modulation is one of the candidate technologies for future
generation broadband wireless networks. Provided that the system parameters are accurately
dimensioned, OFDM transmissions are not affected by intersymbol interference (ISI) even in highly
dispersive channels. Moreover, OFDM can effectively exploit the channel frequency diversity
\cite{Keller&Hanzo,Cheng1993} by dynamically adapting power and modulation format on all
subcarriers. Orthogonal frequency multiple access (OFDMA) is the multiple access scheme based on OFDM:
 each user is allocated a different subset of orthogonal subcarriers. When the transmitter possesses full
knowledge of channel state information, the subcarriers can be
allocated according certain optimality criterion to increase the
overall spectral efficiency, exploiting the so-called
\emph{multiuser diversity}. Resource allocation is one of the most
efficient techniques to increase the performance of multicarrier
systems. In fact, propagation channels are independent for each user
and thus the sub-carriers that are in a deep fade for one user may
be good ones for another. Many resource allocation algorithms have
been designed for taking advantage of both the frequency selective
nature of the channel and the multi-user diversity. In most cases dynamic resource allocation has been formulated with the goal of either
minimizing the transmitted power with a rate constraint \cite{Kivanc,Kim} or  maximizing the overall rate with a power constraint
\cite{Rhee,Jang}.
\par
In this paper, starting from the formulation of resource allocation problem as a minimization problem,
we propose a novel distributed resource allocation scheme for the up-link of a cellular
multi-carrier system based on the message passing (MP) algorithm.  MP algorithms have gained their momentum in the last years owing to their broad usage in LDPC and turbo channel decoding applications \cite{Rich}. In this setting, messages represent
probabilities or beliefs \footnote{the algorithm is also known as the belief propagation algorithm} ,
which are exchanged with the goal of achieving an optimal bit decisions. We will show that resource
allocation may rely on a similar MP procedure: with the goal of
achieving a global optimal assignment, each transmitter iteratively sends and receives
information messages to/from the base station until an allocation decision is taken. The exchanged messages are the solution of small distributed
allocation problems. To reduce the computational load, the MP problems at the terminals follow a
dynamic programming formulation. The advantage of the proposed scheme is that it distributes the
computational effort among all the transmitters in the cell and it does not require the presence of a
central controller for a problem that in its original formulation is NP-hard, as pointed out at the end of Section \ref{sec: sys_model}.
\par
The rest of the paper is organized as follows. In Section \ref{sec:
sys_model} we describe the system model. In Section \ref{sec: mess
pass} we show how message passing can be tailored to solve the
problem of resource allocation in the uplink of a cellular system.
In Section \ref{sec: num res} we present simulation results.
Finally, in Section \ref{sec: concl} we discuss future work and draw
our conclusions.

\section{System model}\label{sec: sys_model}
We focus on the problem of channel allocation for the uplink of an OFDMA system. The overall frequency
bandwidth is divided into orthogonal sub-carriers and, to reduce allocation complexity, we group
sets of adjacent subcarriers into $F$ \emph{subchannels}. As long as the bandwidth spanned by a subchannel
is smaller than the channel coherence bandwidth, the channel spectrum can be approximated as flat in
the subchannel. Thus, we can assume that the choice of performing resource allocation on subchannels
rather than on subcarriers causes almost no loss in diversity.

Allocation is performed with the goal of minimizing the overall
transmitted power subject to rate constraints per user. Due to
practical considerations, we consider only a limited set
$\mathcal{Q}=\left\{0,\hdots,Q\right\}$ of possible transmission
formats. A given transmission format $q$ corresponds to the usage of
a certain error correction code and symbol modulation that leads to
a spectral efficiency $\eta_q$:  a user employing format $q$ on a
certain subchannel transmits with rate $R=B\eta_q$, $B$ being the
bandwidth of each subchannel. The spectral efficiency associated
with format $q=0$ is $\eta_0=0$, i.e. no transmission at all. The
target SNR to achieve the spectral efficiency
$\eta_q=\log_2\left(1+SNR(q)\right)$ is $SNR(q)=2^{\eta_q} - 1$. Let
$\mathcal{F}$ be the set containing the $F$ available subchannnels.
Given the format $q$, the power $P_{n,f}(q)$ necessary to user $n$
to transmit on subchannel $f$ is computed as
\begin{equation}
P_{n,f}(q)=SNR(q)\frac{BN_0}{\left\vert H_{n,f}\right\vert^{2}}
\end{equation}
where $H_{n,f}$ is the channel gain between user $n$ and the BS on the $f$-th link and $N_0$ is the
power spectral density of the zero-mean thermal noise. Channel assignment is exclusive: each
subchannel can be assigned to only one user and with a just a single format.\par
Our resource allocation problem is a constrained minimization
problem in the vector $\mathbf{x}=\left[x_{1,1},\hdots,x_{N,F}\right]$, where
the variable $x_{n,f}\in \mathcal{Q}$ indicates the modulation
format of user $n$ on subchannel $f$. The allocation problem has the following general form
\begin{gather}
\text{minimize } f_0(\mathbf{x}) \label{eq:O}
\\
\text{subject to} \nonumber\\
d_f(\mathbf{x})\leq 1 \qquad f\in \mathcal{F}  \tag{C1} \label{eq:OC1}\\
h_n(\mathbf{x}) \ge b_n \qquad n=1,\dots ,N \tag{C2}
\label{eq:OC2}\\
g_n(\mathbf{x}) \le P_{max,n} \qquad n=1,\dots ,N \tag{C3}
\label{eq:OC3}
\end{gather}
Here the objective function $f_0:\mathcal{D} \to \mathbb{R}_+$ is the
cost in terms of overall power of the allocation $\mathbf{x}$:
\begin{equation}
f_0(\mathbf{x})=\sum\limits_{n=1}^N\sum\limits_{f\in \mathcal{ F}}P_{n,f}(x_{n,f})
\end{equation}
the domain $\mathcal{D}=\mathcal{Q}^{NF}$ is the set of all possible transmission formats on all
subchannels for all users. The inequality constraints functions  $d_f:\mathcal{D} \to \mathbb{R}_+$
represent the condition of exclusive allocation for all subchannels
\begin{equation}
d_f(\mathbf{x})=\sum\limits_{n=1}^N\mathcal{I}(x_{n,f})
\end{equation}
where $\mathcal{I}(x_{n,f})$ is 1 if $1\leq x_{n,f} \leq Q$ and 0 otherwise. The constraints functions
$h_n:\mathcal{D} \to \mathbb{R}_+$ enforce that each user transmits at least with rate $b_n$
\begin{equation}
h_n(\mathbf{x})=\sum\limits_{f \in \mathcal{F}} B \eta_{x_{n,f}}
\end{equation}
The constraints functions $g_n:\mathcal{D} \to \mathbb{R}_+$ enforce
that each user does not exceed its maximum transmitting
power $P_{max,n}$
\begin{equation}
g_n(\mathbf{x})=\sum\limits_{f \in \mathcal{F}}P_{n,f}(x_{n,f})
\end{equation}
%

\par
The radio resource allocation problem introduced above can be shown
to be NP-hard by a straightforward reduction from the NP-hard
problem {\em Multiprocessor Scheduling} \cite{gj}, even when only
one single transmission format is considered.

\section{Resource Allocation via Message Passing}\label{sec: mess pass}
In the following, we formulate the allocation problem in such a way
that can be solved with a message passing technique (MP). The
advantage of MP is that the computation load is distributed among
the various nodes by locally  passing simple messages among simple
processors whose operations lead, after some time, to the solution
of a global problem.

First of all, to simplify the allocation task we assume that each user selects a subset of all
available subchannels. Let $\mathcal{P}_n \subset \mathcal{F}$ be the subset of cardinality $P < F$ of
subchannels that can be allocated to user $n$.  In other terms, we assume that $x_{n,f}$ may be
different from zero only if $f\in \mathcal{P}_n$. As for the choice of the subchannels in
$\mathcal{P}_n$, we make the natural assumption that they represent the $P$ best subchannels for user
$n$, i.e. $\mathcal{P}_n =\left\{f \in \mathcal{F}:\left|H_{n,f}\right| \text{is one of the } P \text{
largest values for user } n\right\}$. Each user may pre-compute its subset of channels before the
resource allocation algorithm is initiated \footnote{We assume perfect channel state estimation
between each user and its serving BS.}. \\
For our scope, it is convenient to interpret the resource allocation problem as a minimum cost
problem, where the unfulfillment of constraints in (\ref{eq:O}) gives an infinite cost. Thus, we take
care of the constraints \ref{eq:OC1} by introducing the cost function $C(f)$ ($f \in \mathcal{F}$),
which is $0$ if the exclusive requirement on subchannel $f$ is fulfilled and $\infty$ otherwise
\begin{equation}
C(f) = \left\{\begin{array}{cc}
                       0 & ~ ~ \text{if}
                       \sum\limits_{n \in \mathcal{N}(f)} \mathcal{I}(x_{n,f}) \le 1 \\
                      \infty &  ~ ~  \text{otherwise} \\
                     \end{array}\right.
 \label{eq9}
\end{equation}
 where $\mathcal{N}(f)$ is the subset of users that might use subcarrier $f$,
 i.e.
$\mathcal{N}(f) =\left\{n: f\in \mathcal{P}_n\right\}$.
The constraints \ref{eq:OC2} and \ref{eq:OC3} are dealt by introducing the set of functions
$W(n)$ ($n = 1,\hdots,N$), defined as:
\begin{equation}
W(n) = \left\{\begin{array}{cc}
                       \sum\limits_{f \in \mathcal{P}_n}
                       P_{n,f}\left(x_{n,f}\right)&\text{if}
                      \sum\limits_{f \in \mathcal{P}_n} B \eta_{x_{n,f}} \ge b_n \\
                        & \sum\limits_{f \in \mathcal{P}_n}P_{n,f}(x_{n,f})\le P_{max,n}\\

                       \infty &  ~ ~  \text{otherwise} \\
                     \end{array}\right.
 \label{eq10}
\end{equation}
Despite notation complexity, the meaning of $(\ref{eq10})$ is
straightforward: $W(n)$ is the power transmitted by user $n$
if power and rate constraints for user $n$ are fulfilled, and $\infty$ otherwise.\\
Given the above, it is straightforward to rewrite the resource
allocation problem in (\ref{eq:O}) as:
\begin{equation}
\hat{\mathbf{x}}=\mathop{arg~min}\limits_{\mathbf{x}}\left(\sum\limits_{f \in \mathcal{F}}C(f)
+\sum\limits_{n=1}^{N}W(n)\right).
\label{eq11}
\end{equation}
Since the goal is to get a distributed solution for the above minimization problem, we focus on a
single variable, e.g., $x_{n,f}$, and rewrite the same problem in a form suited for MP implementation
as:
\begin{equation}
{\hat{x}_{n,f}} = {\mathop{arg ~
min}\limits_{{x}_{n,f}}}\left[\mathop{ min}\limits_{ \bar{x}_{n,f}}
\left(\sum\limits_{f \in \mathcal{F}}C(f)
+\sum\limits_{n=1}^{N}W(n)\right)\right] \label{eq12}
\end{equation}
where notation $\mathop{ min}\limits_{ \bar{x}_{n,f}}$ denote the
minimum over all variables $\mathbf{x}$ except ${x}_{n,f}$.

\subsection{MP implementation}

\begin{figure}[ptb]
\begin{center}
\includegraphics[width = 7.5cm]{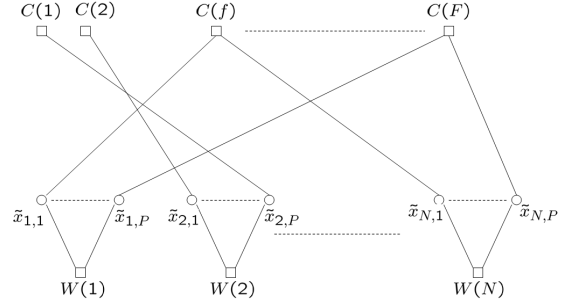}
\caption{Factor graph for RRM. For ease of representation, we denote by $\tilde{x}_{n,p}$
 $(p=1,\hdots,P)$ the transmit format for user $n$
on the p-th ordered element of $\mathcal{P}_n$. }%
\label{Fig4}
\end{center}
\end{figure}

The MP algorithm has been broadly used in the last years in channel coding applications. In
particular, when dealing with bitwise MAP channel decoding, MP finds an optimum solution for the
sum-product problem, provided that the correspondent factor graph is a tree \cite{David}. The MP
algorithm for the sum-product problem derives by the distributive law, i.e., by the property $\sum
\prod = \prod \sum$. However, since $\min \sum = \sum \min$, the same property still holds for min-sum
problems, where minimization replaces addition in the original formulation\footnote{See \cite{Rich},
\cite{David} for a detailed description of MP algorithm for the sum-product problem.} and addition
replaces multiplication. By exploiting such a formal equivalence, it is straightforward to adapt the
MP algorithm to the min-sum problem (\ref{eq12}). To elaborate, let associate with problem
(\ref{eq12}) a factor graph, where variables ${x}_{i,p}$ are circular nodes and functions $C(f)$ and
$W(n)$ are square nodes. Variable nodes are connected with function nodes by an edge if and only if
the variable appears in the function, i.e. ${x}_{n,p}$ is connected to the $P$ functions $C(\ell)$
with $\ell \in \mathcal{P}_n$ and to $W(n)$. The factor graph for (\ref{eq12}) is depicted in Fig.
\ref{Fig4} where we denote by $\tilde{x}_{n,p}$
 $(p=1,\hdots,P)$ the transmit format for user $n$
on the p-th ordered element of $\mathcal{P}_n$. Following a MP strategy, variable and function nodes
exchange messages along their connecting edges until variable nodes can decide on the value of
${\tilde{x}}_{i,p}$.
\par
Let now assume that the factor graph is a
single tree, i.e., a connected graph where there is an unique path
to connect two nodes. In this case, the implementation of the MP approach is straightforward.
Let firstly introduce messages as $(Q+1)$-dimensional vectors, denoted by
$\mathbf{m} = \left\{m(0),m(1),\ldots,m(Q)\right\}$. In particular,
denote by $\mathbf{m}^{(CV)}_{n,f} / \mathbf{m}^{(VC)}_{n,f}$ messages exchanged
between the $C$ function nodes and the connected variable nodes, and by
$\mathbf{m}^{(WV)}_{n,f} / \mathbf{m}^{(VW)}_{n,f}$ messages exchanged between the $W$
function nodes and the connected variable nodes.

As in the classical sum-product scenario, message passing starts at the leaf nodes,
i.e., those nodes which have only one connecting edge. In
particular, each variable leaf node passes an all-zero message to
its adjacent function node, whilst each function leaf node passes
the value of the function to its adjacent node. After initialization at leaf nodes, for every node we can compute
the outgoing message as soon as all incoming messages along all
other connected nodes are received.

As far as variable nodes are of concern, the outgoing message sent over an edge is simply evaluated by
summing all messages received from the other edges. With regard to function nodes, let first consider
the $C(f)$ nodes and focus on generic subchannel $\ell$. The square node corresponding to $C(\ell)$ is
connected to all the variable nodes $x_{n,\ell}$ with $n \in \mathcal{N}(\ell)$. If we consider
without loss of generality the message to be delivered to $x_{j,\ell}$ $\left(j \in \mathcal{N}(\ell)
\right)$, the $q$-th element of the  return message is the solution of the following minimization
problem:
\begin{equation}
\begin{array}{c}
{m}^{(CV)}_{j,\ell}(q) = \min\sum\limits_{n\in\mathcal{N}(\ell),n\ne j}m^{(VC)}_{n,\ell}(x_{n,\ell}) \\
 \text{subject to}\\
 \sum\limits_{n\in\mathcal{N}(\ell),n\ne j}\mathcal{I}(x_{n,\ell}) + \mathcal{I}(q) \le 1
\end{array}
 \label{eq13}
\end{equation}

In a similar way, we can rewrite message passing rule for $W(n)$
nodes. Let focus on generic user $u$, the square node
corresponding to $W(u)$ is connected to the variable nodes
$x_{u,f}$ with $f \in \mathcal{P}_u$. If we consider without loss
of generality the message to be delivered to $x_{u,\nu}$
$\left(\nu \in \mathcal{P}_u \right)$, the $q$-th element of the
return message is the solution of the following  minimization
problem:
\begin{equation}\label{eq14}
 \begin{array}{c}
  {m}^{(WV)}_{u,\nu}(q) =\min
                       \sum\limits_{f\in\mathcal{P}_u}P_{u,f}\left(x_{u,f}\right) + {m}^{(VW)}_{u,f}(x_{u,f}) \\
\text{subject to}\\
\sum\limits_{f\in\mathcal{P}_u,f\ne \nu} B \eta_{x_{u,f}} + B \eta_q \ge b_u\\
\sum\limits_{f\in\mathcal{P}_u,f\ne \nu} P_{u,f}(x_{u,f}) +P_{u,\nu}(q) \le P_{max,u}
\end{array}
\end{equation}
When a message has been sent in both directions along every edge the algorithm stops. It is worth
noting that in the considered OFDMA cellular scenario the $W(n)$ function node and its connected variable
nodes are located at the $n$-th user, while all $C(f)$ function nodes are located at the BS. Hence,
sending messages from variable nodes to $C(f)$ function nodes and vice-versa requires actual
transmission on the radio channel. Instead, message exchange between variable nodes and $W(n)$ function
nodes is performed locally at the users' terminals, without any transmission.\\
The solution of Problem \eqref{eq14} requires by far the largest computational effort, since it calls
for an exhaustive search over all possible combinations of transmission formats. Thus, in the
following we present a new formulation of \eqref{eq14} to find the optimal solution with limited
complexity.

\subsection{A Dynamic programming algorithm}
Given a user $u$, Problem \eqref{eq14} basically consists in finding a set of subchannels $f
\in\mathcal{P}_u$, and for each selected subchannel the related transmission
format to use by $u$, so that a given function is minimized. Such problem can formulated as an Integer
Linear Programming (ILP) problem introducing binary variables $y_{f,h}$ equal to 1 if the user
transmits on the subchannel $f$ with the format $h$, and 0 otherwise. In a general form, such a
problem can be rewritten as
\begin{eqnarray}\label{prob:single_user}
\min \sum\limits_{f,h} c_{f,h} y_{f,h}\nonumber\\
\sum\limits_{f,h} B\eta_{h} y_{f,h} &\geq& \beta\nonumber\\
\sum\limits_{f,h, f \not=\nu} P_{u,f}(h) y_{f,h} &\leq& \alpha\\
\sum\limits_{h} y_{f,h} &\leq& 1 \; \; f \in \mathcal{P}_u \nonumber\\
y_{f,h} \in \{0,1\}\nonumber
 \end{eqnarray}
where the cost $c_{f,h}$ is the cost for user $u$ of transmitting with format $h$ on subchannel $f$
(i.e., $c_{f,h}=P_{u,f}\left(h\right) + {m}^{(VW)}_{u,f}(h)$), $\beta=b_u-B\eta_q$, $\alpha=
P_{max,u}-P_{u,\nu}(q)$. As in \eqref{eq14}, the first two constraints correspond to the requirements
on the bit-rate $b_u$ and on the maximal transmission power $P_{max,u}$, respectively. The subsequent
$P$ constraints impose that at most one format is selected for each subchannel $f$. Note that, $b_u$
is limited from above by $B\eta_QP$, and such a value is obtained when user $u$ transmits with format
$Q$ on all the subchannels in $P_u$. Assuming that all admissible formats are multiple integer of a
given spectral efficiency $\tilde\eta$, ie $\eta_h=h\tilde\eta$ ($h=0,\hdots,Q$), we can divide all
terms of the first constraint of Problem \eqref{prob:single_user} by $B\tilde\eta$ to obtain the
equivalent constraint
\begin{equation}\label{constr:beta}
\sum\limits_{f,h} h \ y_{f,h} \geq {\frac{\beta} {B\tilde\eta}}
\end{equation}
where ${\frac{\beta} {B\tilde\eta}}$ is limited from above by $QP$. Observe that, the coefficients $h$
in the left-hand side of constraint \eqref{constr:beta} are integer values. Hence, since the left-hand
side of \eqref{constr:beta} is
 integer, for any choice of variables $y_{f,h} \in
\{0,1\}$, ${\frac{\beta} {B\tilde\eta}}$ can be rounded to $\lfloor {\frac{\beta} {B\tilde\eta}}
\rfloor$. Moreover, we may assume that values $P_{u,f}(h)$ and $\alpha$ are integer (e.g., by
multiplying all terms of the second constraint of Problem \eqref{prob:single_user} by a suitable large
number).

In the following, we show that Problem \eqref{prob:single_user} can be solved by a {\em dynamic
programming} approach \cite{Bellman}. Let $z_p(d,k)$ be the optimal solution value of Problem
\eqref{prob:single_user} defined on the first $p$ subchannels, with a "bit-rate" of $\lfloor
{\frac{\beta} {B\tilde\eta}} \rfloor = d$ and a restricted maximal transmission power of $k$. We
assume that $z_p(d,k)=+\infty$ if no feasible solution exists.
 Initially we set $z_0(0,k)=0$ and $z_0(d,k)=+\infty$ for all
$d=1,\ldots,QP$ and $k=0,1,\ldots,\alpha$. To compute $z_p(d,k)$, we
can use the recursion \eqref{formuladynprog},
\begin{figure*}[t]
\begin{equation}\label{formuladynprog}
z_p(d,k)= \min \left \{ \begin{array}{ll}
                 z_{p-1}(d,k) +c_{p,0} & (\hbox{subchannel} \ p \;\hbox{is not  used by the user})\\
                 z_{p-1}(d-1,k-P_{u,p}(1)) +c_{p,1} & \hbox{if} \ d-1 \geq 0 \ \hbox{and}\ k-P_{u,p}(1)\geq 0  \ ( p \;\hbox{is used with format} \ 1)\\
                 \ldots       &\\
                z_{p-1}(d-Q,k-P_{u,p}(Q)) + c_{p,Q} & \hbox{if} \ d-Q \geq 0 \ \hbox{and}\ k-P_{u,p}(Q)\geq 0  \ ( p \;\hbox{is used with format} \
                Q)
               \end{array}
        \right.
    \end{equation}
\end{figure*}
where we assume that the minimum operator returns $+\infty$ if we are minimizing over an empty set. An
optimal solution of Problem \eqref{prob:single_user} can be found computing $z_{P}(QP,\alpha)$, and
choosing the minimum of $z_{P}(j,\alpha)$ for $j=\lfloor {\frac{\beta} {B\tilde\eta}} \rfloor,\ldots,
QP$. The formula \eqref{formuladynprog} requires the comparison of $Q$ terms, and, hence, the optimal
solution value of Problem \eqref{prob:single_user} can be found in $O(P^2Q^2\alpha)$ operations, only
pseudopolynomial, since $\alpha$ depends on the input data (i.e,
 $P_{max,u}$ and $P_{u,\nu}(q)$). Observe that, if no
requirement is given on the maximum transmission power used by each
user, i.e., if the constraint on $\alpha$ can be relaxed, Problem
\eqref{prob:single_user} can be solved in $O(P^2Q^2)$ operations,
polynomial in the number of subchannels in $\mathcal{P}_u$ and transmission
formats.

\subsection{MP scheduling and Peeling procedure}

As in traditional MP approach for the sum-product problem, if the
factor graph is a tree there is a natural schedule for MP given by
starting at the leaf nodes and sending a message once all incoming
messages required for the computation have arrived \cite{Abr1}.
Unfortunately, in general the graph which represents minimization
problem (\ref{eq12}) is not a tree (e.g., in Fig. \ref{Fig4} we have
a cycle given by the path
$\tilde{x}_{1,1},C(f),\tilde{x}_{N,1},\tilde{x}_{N,P},C(F),\tilde{x}_{1,P},\tilde{x}_{1,1}$).
In this case, to completely define the algorithm for a generic
factor graph we need to specify a schedule. It is worth noting that,
even if message passing in the presence of cycles is strictly
suboptimal, the solution found by means of iterative approaches is
in most cases very close to the optimum (e.g., in the case of
bitwise MAP decoding of linear block codes) \cite{Abr2},\cite{Abr3}.\\
Iterative MP starts at variable nodes, which send an all zero message $\mathbf{m}_{n,f}^{(VW)} = 0$ to
their adjacent $W(n)$ function nodes and then the algorithm proceeds in iterations. The pseudocode of
Algorithm \ref{algMP} illustrates the iterative MP algorithm for a generic user $n$. After $I$
iterations, each user peels off all variable nodes $x_{n,f}^{(I)}> 0$. All these variable nodes, say
it \emph{fulfilled nodes}, send a message to the BS to communicate that the corresponding subchannels
have been reserved and the BS signals it to all users via the downlink broadcast channel.

\begin{algorithm}
\small \caption{Iterative MP procedure for user $n$} \label{algMP}
\begin{algorithmic}
%
\While{$\sum\limits_{f\in\mathcal{P}_n} B \eta_{x_{n,f}} < b_n$} \State $\mathbf{m}_{n,f}^{(VW)}
\leftarrow 0$ ($f\in\mathcal{P}_n)$ \State send $\mathbf{m}_{n,f}^{(VW)}$ to $W(n)$
($f\in\mathcal{P}_n$)\State \For{$iter=0$ to $I$} \State evaluates $\mathbf{m}_{n,f}^{(WV)}$ according
to (\ref{eq14}) ($f\in\mathcal{P}_n)$ \State $\mathbf{m}_{n,f}^{(VC)} \leftarrow
\mathbf{m}_{n,f}^{(WV)}$ ($f\in\mathcal{P}_n$) \State send $\mathbf{m}_{n,f}^{(VC)}$ to $C(f)$
($f\in\mathcal{P}_n$) \State \While{not received all $\mathbf{m}_{n,f}^{(CV)}$ from $C(f)$
($f\in\mathcal{P}_n$)} \State wait\EndWhile \State $\mathbf{m}_{n,f}^{(VW)} \leftarrow
\mathbf{m}_{n,f}^{(CV)}$ ($f\in\mathcal{P}_n$)\EndFor \State $\mathbf{m}_{n,f} \leftarrow
\mathbf{m}_{n,f}^{(CV)} + \mathbf{m}_{n,f}^{(WV)}$ ($f\in\mathcal{P}_n$) \State $x_{n,f} \leftarrow
\mathop{arg ~ min}\limits_{ q = 0,1,\ldots,Q} \mathbf{m}_{n,f}(q)$ ($f\in\mathcal{P}_n$) \State Sends
a message containing assignments $x_{n,f}$ to the BS

\EndWhile
\end{algorithmic}
\end{algorithm}

Variables corresponding to fulfilled nodes are fixed and do not pass any message anymore. At this
stage, all users evaluate wether they fulfill their rate constraints or not. Those users that satisfy
their constraints stop participating to MP. All other users take part to successive iterations of MP,
after having updated their rate constraints in (\ref{eq14}) on the base of the amount of resources
they have been allocated. Before starting a new cycle of $I$ iterations, each user computes again the
set $\mathcal{P}$ of the best $P$ subchannels among all subchannels which have not been yet assigned
to other users. The process continues until the rate constraint is fulfilled.


\section{Numerical results}\label{sec: num res}
In this section we present the numerical results of the proposed
algorithm. We have considered an hexagonal cell of radius $R=500$ m.
The uplink bandwidth is $W=5$ MHz so the sampling time is
$T_c=200$ ns. Channel attenuation is due to path loss,
proportional to the distance between the BS and the MS, and fading. The path
loss exponent is $\alpha=4$.  We consider a
population of data users with very limited mobility so that the
channel coherence time can be assumed very long. The propagation channel is
frequency-selective Rayleigh fading. The power of the $j$-th path is:
$\sigma_{j}^{2}=\sigma_{h}^{2}\exp\left(-\frac{j}{\sigma_{n}}\right),
(j=1,\dots,N_j)$ where $\sigma_{h}^{2}$ is a normalization factor
chosen such that the average power of the channel is normalized to
the value of the path loss, $\sigma_{n}=\sigma_{\tau}/T_c$ is the
normalized delay spread with $\sigma_{\tau}=0.5~\mu s$ and
$N_j=\lfloor 3 \sigma_{n} \rfloor$ is the number of paths taken into
an account.
\par
\begin{figure}[ptb]
\begin{center}
\includegraphics[width = 7.8cm]{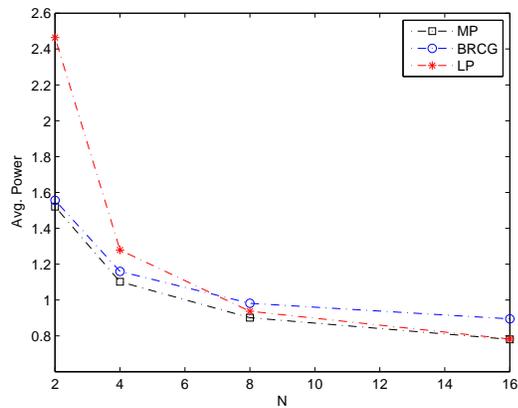}
\caption{Average power consumption versus number of users. }%
\label{Pl1}
\end{center}
\end{figure}
\begin{figure}[ptb]
\begin{center}
\includegraphics[width = 7.88cm]{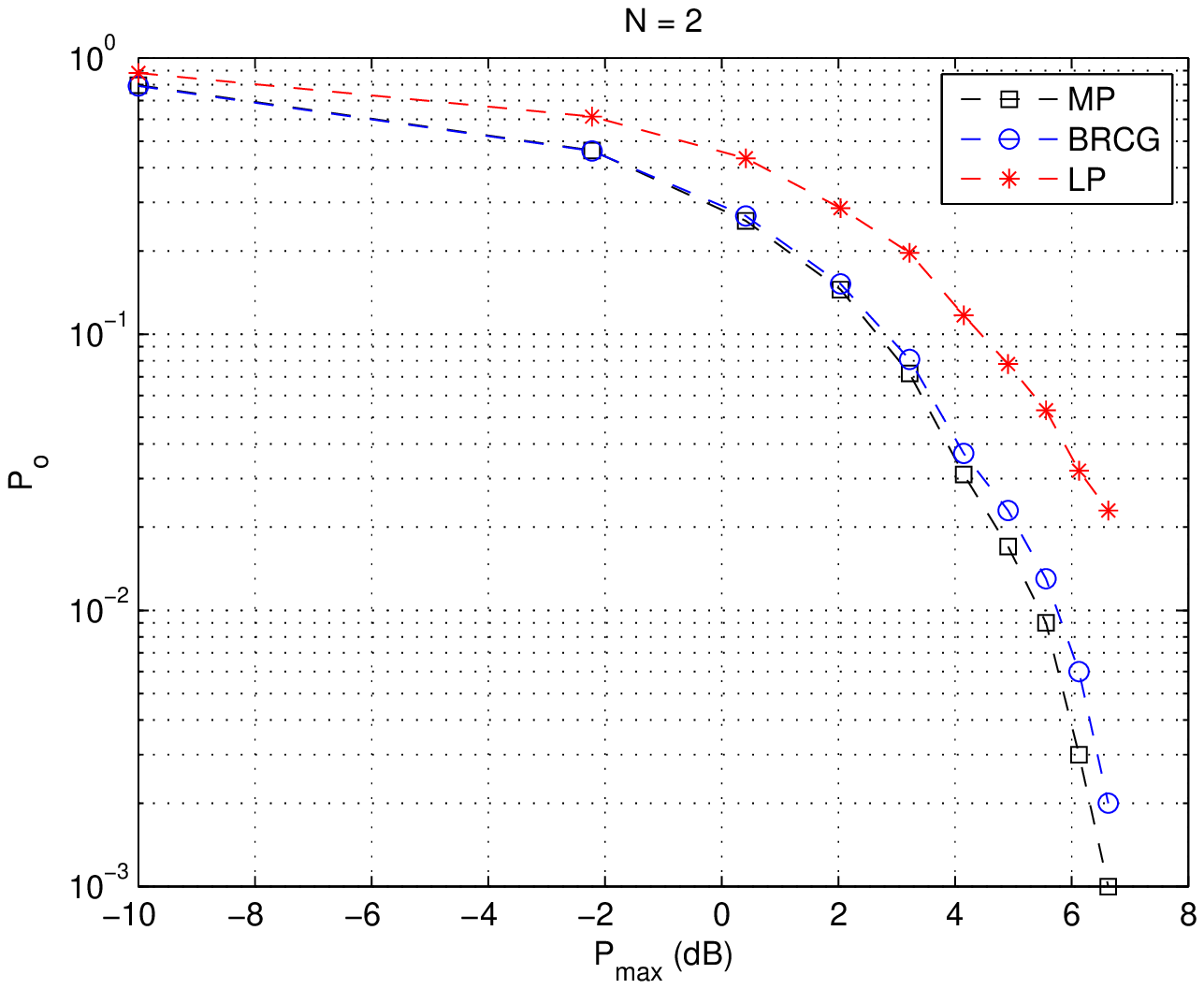}
\caption{Outage probability versus maximum transmitting power for $N = 2$. }%
\label{Pl2}
\end{center}
\end{figure}
\begin{figure}[ptb]
\begin{center}
\includegraphics[width = 7.8cm]{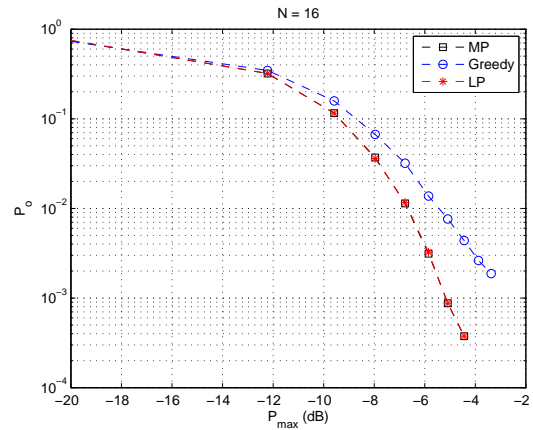}
\caption{Outage probability versus maximum transmitting power for $N = 16$. }%
\label{Pl5}
\end{center}
\end{figure}

The available bandwidth  $W$ is divided in $F=32$ subchannels and there are $N$ active users at one
time. We assume that all users request the same rate i.e., $b_n=b_0$, ($n=1,\hdots,N$), so that
$b_0=W\eta_{avg}/N$,  where $\eta_{avg}$ is the  average spectral efficiency in the cell.  The results
shown in the following have been obtained by setting $\eta_{avg}=1$ b/s/Hz and averaging on 500
channel realizations. We compare the performance of the proposed MP algorithm with two other resource
allocation strategies:
\begin{enumerate}
\item The heuristic algorithm presented in \cite{Kivanc} that we have indicated with the acronym BRCG
(Babs +  RCG)  that  solves the problem \eqref{eq:O} by dividing it in three subproblems: 1) Decide
the number of subcarriers each user gets based on rate requirements and the users average channel gain
(bandwidth assignment based on SNR, BABS); 2) Select which subcarriers to allocate to each user
according a greedy strategy (rate craving greedy, RCG); 3) Set the modulation for each subcarriers by
employing a single-user bit loading technique.

\item A linear programming (LP) implementation of the allocation problem \eqref{eq:O} formulated as in
\cite{Kim} where the rate constraints are translated into a number of subchannels to assign to each
user. In our implementation, we set a unique transmission format for all users on all subchannels, so
that each user is assigned the same number of subchannels $F/N$ and transmits with spectral efficiency
$\eta=\eta_{avg}$. By doing so,  we neglect on purpose the impact of frequency diversity to focus only
on the impact of multi-user diversity on the allocation performance.
\end{enumerate}
As far as MP and BRCG are of concern, we set $Q = 4$, i.e., we consider four different transmission
formats. Fig. \ref{Pl1} shows the average total transmitting power for different number of users. As
far as MP parameters are of concern, we set $P = 4,8,12,24$ for $N = 2,4,8,16$, respectively.
Moreover, since both BRCG and LP do not take into account any constraint on transmitting power, we set
maximum transmitting power to $+\infty$ in (\ref{eq14}). Note that the proposed MP algorithm requires
the minimum average power in all cases. In particular, for small values of $N$, MP and BRCG clearly
outperform LP, whilst for high values of $N$, MP and LP outperform BRCG. With few users, i.e., for
small values of $N$, each user is assigned a great number of subchannels. In this case, the use of
multiple transmission formats in MP and BRCG algorithms allows the transmitter to concentrate the
power on the \emph{best} channels while turning off the \emph{worst} ones. Although designed to
achieve an optimal global subchannel allocation,  the LP scheme shows poor performance since it is
forced to use the same transmission format over all assigned subchannels. On the other hand,
increasing the number of users reduces the number of channels allocated and each user is assigned only
'good' channels, thus exploiting the so-called multi-user diversity. Fig. \ref{Pl1} shows that for all
three algorithms increasing the number of users determines a reduction of the average transmission
power. However, the BRCG algorithm can not fully exploit multi-user diversity, since greedy channel
assignments are sub-optimal and, even using several transmission formats,  is outperformed by the LP
scheme
already for $N \ge 8$.  \\
Similar considerations can be drawn when considering the outage probability curves, i.e., Figs.
\ref{Pl2}-\ref{Pl5}. The outage probability $P_o$ is evaluated by specifying a maximum allowable
transmitting power $P_{max}$ for each user. Such a maximum power is included in (\ref{eq14}), so that
outage events in the MP case occur when the iterative MP algorithm is not able to provide a feasible
solution for all users. Differently, since in the LP and BRCG approaches we have not included
constraints on the maximum transmitting power, outage events occur when, after the allocation, the
power transmitted by a user exceeds $P_{max}$. Figs. \ref{Pl2}-\ref{Pl5} show $P_o$ as a function of
$P_{max}$ for $N=2$ and $N=16$. The proposed MP scheme achieves the lower outage probability in all
cases, thus confirming that it allows to perform an optimal subchannel assignment and to profitably
exploit both frequency and multi-user diversity. Furthermore, although the computational complexity of
MP depends on the number of iterations, at each iteration, the computation is naturally distributed
among the transmitters, which have to solve low-complexity and in practice small problems.

\section{Conclusion}\label{sec: concl}
We have  proposed a novel distributed resource allocation scheme for
the up-link of a cellular multi-carrier system based on the message
passing (MP) algorithm. Resource allocation may rely on a similar MP
procedure: with the goal of achieving a global optimal assignment,
each transmitter iteratively sends and receives information messages
to/from the base station until an allocation decision is taken. The
exchanged messages are the solution of small distributed allocation
problems. To reduce the computational load, the MP problems at the
terminals follow a dynamic programming formulation. Hence, even if
the computational complexity of MP depends on the number of
iterations, at each iteration, the computation is naturally
distributed among the transmitters, which have to solve
low-complexity and in practice small problems. Numerical results
show that the proposed approach is an excellent solution to the
resource allocation problem for a single-cell multi-carrier system.
Moreover, the distributed nature of the proposed strategy make it
naturally suitable for larger scale resource allocation problems,
such as global resource optimization in multi-cell OFMA systems.
\bibliographystyle{IEEE}

\end{document}